\documentclass[preprint]{JHEP3}

\usepackage{epsfig,multicol}

\newcommand{\be}{\begin{equation}}
\newcommand{\ee}{\end{equation}}
\newcommand{\bear}{\begin{eqnarray}}
\newcommand{\eear}{\end{eqnarray}} \newcommand{\ba}{\begin{array}}
\newcommand{\ea}{\end{array}}

\preprint{FERMILAB-PUB-10-005-A}

\title{A Faraway Quasar in the Direction of the Highest Energy Auger 
Event} 

\author{Ivone F.M. Albuquerque\\ 
Center for Particle Astrophysics, Fermi National Accelerator 
Laboratory, Batavia, IL, 60510 and Instituto de F\'isica, Universidade de 
S\~ao Paulo, S\~ao Paulo, Brazil\\
\email{ifreire@fnal.gov}}
\author{Aaron Chou\\ 
Center for Particle Astrophysics, Fermi National Accelerator Laboratory,
Batavia, IL, 60510\\
\email{achou@fnal.gov}}

\abstract{
The highest energy cosmic ray event reported by the Auger Observatory has an
energy of 148 EeV. It does not correlate with any nearby (z$<$0.024) 
object capable of
originating such a high energy event. Intrigued by the fact that the highest energy
event ever recorded (by the Fly's Eye collaboration) points to a faraway quasar
with very high radio luminosity and large Faraday rotation measurement, we 
have searched for a similar source for the Auger event. We find that the Auger 
highest energy event points to a quasar with similar
characteristics to the one correlated to the Fly's Eye event. 
We also find the same kind of correlation for one of the highest energy AGASA 
events. We conclude that so far these types of quasars are the best source candidates for both
Auger and Fly's Eye highest energy events. We discuss a few exotic candidates 
that could reach us from gigaparsec distances. 
}


\begin{document}

\section{\label{introduction} Introduction}
Observatories such as the Pierre Auger and the High Resolution 
(HiRes) Fly's Eye have recently shed light on some of the open questions 
regarding the high energy end of the cosmic ray spectrum. However, much is yet
to be revealed. Among the open questions are the origin of 
ultra high energy (above $\sim 6 \times 10^{19}$~eV) cosmic rays (UHECR), the acceleration mechanism involved in their production, and their
composition.  

As a first step in the search for the astrophysical sources of UHECR, the Auger collaboration has observed an anisotropy of the
UHECR arrival direction  at 99\% CL \cite{augani}, via a correlation of cosmic ray arrival directions with nearby Active Galactic Nuclei
(AGNs).  The correlation with this anisotropic pattern in the sky does not imply that the AGN are necessarily the sources of UHECR, but rather that the cosmic rays tend to come from where the large scale structure of matter density is distributed.  It suggests that the deflection of cosmic ray trajectories by galactic and extragalactic magnetic fields may be small enough that cosmic ray astronomy becomes possible.  To add to the puzzle, a similar correlation with nearby AGN has not been seen in the HiRes data \cite{hragn}.  


Both experiments have observed \cite{hrgzk,augzk} the expected Greisen-Zatsepin-Kuzmin (GZK) suppression \cite{gzk} in the cosmic ray 
spectrum due to resonant scattering on the cosmic microwave background photons. Furthermore, the Auger correlation with nearby AGN is compatible with the expectation of the resulting GZK particle horizon. 


Here we focus our research on the highest energy events, those with energy above 
$10^{20}$~eV.  
Naively, one would expect that the highest energy UHECRs would be the least deflected by magnetic fields, and hence would be the events most likely to point back to their astrophysical sources.   Paradoxically, there are no known plausible sources within the GZK sphere for the highest energy events, leading some to hypothesize that UHECR are produced by nearby, energetic, transient sources such as gamma ray bursts~\cite{Waxman:2004ez,Lemoine:2009pw} or AGN flares~\cite{Farrar:2008ex,Waxman:2008bj}.  These sources would not necessarily have large photon fluxes observed concurrently with a UHE hadronic cosmic ray flux. 

These events might also be heavy nuclei being deflected 
by the Galactic magnetic field. As it is easier to accelerate heavy nuclei than protons, one
can analyze the possibility of heavy nuclei acceleration. 

Another exciting possibility is that new, exotic physics may be responsible for the unimpeded propagation of UHECRs over cosmological distances.  Although the presence of the GZK feature is compatible with a
standard particle composition of UHECR, it does not rule out the possibility of a small fractional flux of an exotic component.  Such an exotic flux may indeed be able to penetrate through the opaque wall of background photons and travel far larger distances without attenuation.

Inspired by a thorough search \cite{som} for the source of the highest energy event so far
detected (by the Fly's Eye collaboration \cite{fly}), we search for a 
possible source of the Auger highest energy event.
This event does not correlate with any nearby AGN from the V\'{e}ron-Cetty 
and V\'{e}ron catalog \cite{vc} used in the Auger correlation studies nor
with any nearby source capable of producing such an energetic event.  This lack of correlation is particularly puzzling because the event points back to an apparently well-covered region of the sky between the Virgo and Centaurus superclusters, where the VC catalog would be expected to be complete for the highest luminosity, nearby AGN.  
Instead, we find that the most energetic event so far detected by the Auger
collaboration, points to a faraway quasar, well beyond the GZK radius. 
This quasar has the same characteristics as the one which
correlates with the Fly's Eye event.

We also analyze the arrival direction of AGASA's 11 events above 
$10^{20}$~eV~\cite{agasa}. It should be noted however that AGASA's energy 
scale is systematically 30\% higher than the  
HiRes/Fly's Eye energy scale~\cite{demarco} while Auger's energy scale is around 15\% lower.  Unlike the Auger and HiRes measured UHECR spectra, the AGASA spectrum at high energies exhibits no GZK suppression; this may be indicative of poorer energy resolution at these energies.  The HiRes collaboration has not yet published their event list containing
the energy and direction information together. For this reason, it
is not possible for us to analyze their results as we did for Auger and AGASA.  However, it is intriguing that HiRes has already reported a potential correlation of their stereo events with distant BL Lac objects, well beyond the GZK horizon \cite{Gorb,Abbasi:2005qy}.  This correlation cannot be explained by conventional physics models.


In the next section we will review the search for the source of the Fly's Eye highest
energy event. In section~\ref{sec:aughi} we describe our search for source of the Auger 
highest energy event.  Then follows the analysis for AGASA's events.
In section~\ref{sec:cand} we discuss exotic models which may allow cosmic rays to reach us from such faraway sources.

\section{The highest cosmic ray event ever detected}
\label{sec:fly}

The highest energy
cosmic ray event was detected by the Fly's Eye collaboration
\cite{fly}, which determined the energy to $3.2 \times 10^{20}$~eV.  A thorough search for its source 
was done by
Elbert and Sommers (ES) \cite{som}, who looked for strong radio sources 
with strong magnetic fields and large fluxes of energetic particles in the 
direction of the Fly's Eye event. These sources were targeted since
in order to accelerate particles to ultra
high energies, a high magnetic field moving with high velocity is required
\cite{hillas}. Powerful radio galaxies are therefore good UHECR 
source candidates. Their search was limited to a region within an angular
distance of $10^\circ$ from a two sigma error box centered in the event
arrival direction. 

From the four sources that fell into this region,
the quasar 3C~147 was the only one within one sigma from the event 
arrival direction.
It is also the only one which meets the necessary requirements
for particle acceleration. It is listed in
a catalog of 173 powerful radio luminosity sources 
\cite{herbig}.
Its total radio luminosity is about $10^{45}$~ergs~s$^{-1}$ being among
the 28 brightest ones in this catalog. It is also included in a list of
96 plausible candidates for having very large Faraday rotations (FR)
\cite{hifr}, having one of the top four largest FR \cite{kato}. 
The redshift of this source is however 0.545, which puts it at a distance
of 2 Gpc away from us for current cosmological parameters.

\section{Procedure for Source Correlation}
\label{sec:proc}

In this section we describe our procedure for correlating sources with the arrival direction
of the high energy events. We follow the same procedure adopted by ES in looking
for high luminosity sources in the direction of the Fly's Eye event. 
In order to correlate the HEA to a faraway source, we assume the cosmic ray is
electrically neutral. For this reason, the 
angular cut around the event arrival direction is defined as $1.5 \times \alpha_{exp}$,
where $\alpha_{exp}$ is the experimental angular resolution.

As for the Fly's Eye event, we are not looking for any object in
the direction of the detected event, but for sources with the capability of accelerating
particles to $10^{20}$~eV. It has been shown \cite{martin} that a solid lower bound on 
the source luminosity in order to achieve such an acceleration, is of $10^{45} Z^{-2}$~erg/s, 
where Z is the charge of the cosmic ray nucleus. Our goal is to find sources with
$10^{45}$~erg/s minimum luminosity in the direction of the detected event. 
We should point out however that although this bound is valid for
  many acceleration models including the diffusive or Fermi
  mechanisms, it does not hold for mechanisms like inductive
  acceleration that might occur near central black holes in AGNs. Also
  the requirement is on the magnetic luminosity but as in \cite{martin} we
  use it as a bolometric luminosity limit. 

As a first step we select high luminosity sources according to a minimum flux
density. We use the high luminosity criterion established by Wall \& Peacock~\cite{wallp}, 
which requires a minimum flux density of 2~Jy at a frequency of 2.7~GHz.
This allows us to produce a list of astrophysical objects, which
although broad is a first selection step. Once this list is produced we verify which of these
objects have the minimum required luminosity. 

In order to account for measurement uncertainties, we require a minimum source luminosity of 
$L = 10^{44}$~erg/s. One can roughly translate this to a minimum flux density $S_{\rm min}$ at 
a certain frequency $\nu$ in Hz~\cite{som}:
\be
S_{\rm min} \approx \frac{L H_0^2}{4 \pi \nu c^2 z^2} = \frac{5 \times 10^{9}}{\nu z^2} \label{E:Smin}
\label{eq:smin}
\ee
where $c/H_0$ is the Hubble length, z is the source's redshift, and $S_{\rm min}$ is given in 
Jansky. We use the luminous parameter $R_L = S/S_{min}$ defined by ES, where $S$ is the
source flux density in Jansky and require $R_L > 1$. As an example, the nearby Centaurus A
has $R_L = 10^{-3}$ and M87, $2 \times 10^{-3}$. As we will see the quasar 3C~147 found
in the direction of the Fly's Eye event has $R_L = 6$.
This will reduce the number of objects selected and allow us to require a minimum
angular distance from the event arrival direction and then integrate the observed flux density
to determine the luminosity. The total radio luminosity can be determined as 
\cite{herbig}
\be
L = 4 \, \chi^2 \, (1 + z)^2 \, S_{bol,obs} \, h^{-2}
\label{eq:lum}
\ee
where $\chi$ is the comoving distance and $S_{bol,obs}$ is the observed radio 
spectrum integrated from 10~MHz to 100~GHz. 

Both the first list of high luminosity objects as the flux density at various frequencies 
can be obtained from the NASA/IPAC Extragalactic Database (NED)
catalogue ~\cite{ned}.  For each telescope we look for sources
 within their field of view. Extra requirements are that the sources are away from the galactic
plane ($|b|>10^\circ$) where astrophysical catalogs are incomplete, and
located beyond the GZK radius, at a redshift larger than 0.024. The GZK radius is the maximum 
distance from which a standard particle can travel through the cosmic microwave background 
radiation (CMB) and reach the Earth at ultra high energies. For a proton
this distance is about 100 Mpc. The latter requirement comes from the fact that
searches within the GZK cutoff were already done. The result for a nearby search for the 
Fly's Eye event source is described in section~\ref{sec:fly}; and we will describe the Auger 
search in section~\ref{sec:aughi}. It is worth noting that no high luminosity nearby source 
was found in the direction of these events. 

We now describe how we determine the chance probability that a specific correlation comes from 
an isotropic distribution. The probability {\em p} that the arrival direction of a particular 
event, which comes from an isotropic distribution of sources in the sky, points 
by chance to one specific source in the candidates list is given by:
\be
p = \int_A \varepsilon(\delta) \cos(\delta) d\delta d\alpha_R
\ee
where the integration goes over an area $A$ and $\varepsilon(\delta)$ is the experimental 
exposure function which depends on the declination $\delta$ and the zenith angle $\theta$
and is given by
\be
\varepsilon(\delta) \propto \int_0^2\pi \cos\theta d\alpha_R
\label{eq:exp}
\ee where $\alpha_R$ represents the right ascension and \be \cos\theta = \sin\delta \sin l + \cos\delta \cos l \cos\alpha_R \ee where $l$ represents the observational site latitude. The exposure function is normalized to unity over the experimental field of view of the observatory. The area $A$ is the area in the sky determined by circular regions of radius $\beta$ around each candidate source.  For the Auger and Fly's Eye telescopes, there is no exposure for zenith angles above $60^o$ and the integrand of equation~\ref{eq:exp} is zero in this region. For AGASA there is no exposure for zenith angles above $45^o$.

We  determine {\em p} through a Monte Carlo simulation, where we generate a large number of
sources isotropically distributed over $A$, where they follow the pattern 
determined by the exposure function $\varepsilon(\delta)$. 
We then compute the angular distance between these simulated events and the sources that 
belong to our catalogue. {\em p} is determined considering that the event arrival direction
matches the candidate source direction within an angular distance of $1.5 \alpha_{exp}$.

\subsection{Reanalyzing the Highest Energy Event}

Here we reanalize the the Fly's Eye highest energy event, looking for bright sources in its
direction. We follow each of our procedure steps described above. 
We aim at the same time to check our method and if
no other source -- that could have been discovered after the work of ES -- also matches the
arrival direction of the Fly's Eye event.

One important point in this analysis is that the Fly's Eye telescope exposure is
a complicated function of energy and zenith angle. For this reason our analysis determines
the Fly's Eye exposure in a very crude way. The Fly's Eye angular resolution is asymmetric in 
right ascension and declination, being the later much worse with a value of 6 degrees. To
simplify our search we look for sources within 9 degrees from the event direction. This will
of course make the chance probability of having a correlation from an isotropic distribution
very high. 

We produce our first source list by requiring a minimum flux density of 2~Jy 
at a frequency of 2.7~GHz. This list contains 148 sources. 
The second step is to require a minimum flux density as described by 
$R_L$, which reduces this list to 54 objects. 

Our result is compatible with the one found by ES. We find that the only high luminosity
object that correlates with the Fly's Eye event direction is the Quasar 3C~147. The 
caracteristics of this quasar is described in section~\ref{sec:fly} and also in the
work of ES. The chance probability for this correlation is large, being $p = 0.37$ if
we consider all 54 sources in our final catalalogue.
ES quote that the number of sources from
their catalogue expected to fall in their $2\sigma$ error box centered in the
event direction is 0.24. We integrate equation~\ref{eq:lum} and determine 3C~147 luminosity to 
be $10^{45}$~erg s$^{-1}$, the same as found by ES.

This shows that our procedure is compatible with the one done by ES and we now
use it to analyze both Auger and AGASA's events.

\section{The Auger Highest Energy Event}
\label{sec:aughi}

From the 27 high energy events published by the Auger 
collaboration~\cite{augani}, only one is above $10^{20}$~eV, having an
energy of 148~EeV. Events of these energies are expected to 
have its origin outside of our Galaxy~\cite{hillas}. As ordinary galaxies do not
meet the requirements for accelerating particles to such high energies,
we searched for high luminosity radio sources in the direction of the highest
energy Auger event (HEA). We proceed as described in section \ref{sec:proc}.

We produce our first source selection list requiring a minimum flux density of 2~Jy
at a frequency of 2.7~GHz in the ``advanced all-sky survey'' of
the NED. The equatorial coordinates for the HEA are right ascension 
$\alpha_R = 192.7^\circ$ and declination $\delta = -21^\circ$. 
The Auger exposure~\cite{augani} for events with zenith angle below $60^\circ$ is roughly constant in right ascension and is a function of declinations, being nonzero for $\delta < 24.8^\circ$. 


The NED search results in 184 strong radio luminosity objects. 
We further reduce this list by requiring $S_{\rm min}$ as defined in Equation~\ref{eq:smin}
which is equivalent to $R_L > 1$.
From the 184 NED sources, only 67 meet
this minimum luminosity requirement, among which 57 are quasars and 10 are galaxies.
As described in section~\ref{sec:proc} we look for sources with a redshift $z$ beyond 0.024.
As the Auger optimized search adopted $z < 0.018$,
we also look for sources with a redshift between 0.018 and 0.024. This search is described a bit 
further.

As for these high energies the Auger angular resolution is about $1^\circ$ \cite{augani},
we require that the angular distance between the source and the event is at
maximum $1.5^\circ$. 
From our 67 objects catalog,
only one fell in this angular region. It is the 
quasar (QSO) PKS1245-19, with an angular distance of $1.2^\circ$ from the HEA. The chance 
probability for this to happen from an isotropic distribution is 0.013~\footnote{
Although efforts were made to fix the analysis before looking at the data, 
mistakes were made and the data was processed through two different sets of cuts on b 
and the maximum angular distance between the source and the cosmic ray arrival
direction before the analysis was finalized.  So at most, the reported probability 
would be multiplied by a factor of 3 to account for previous trials.}.

The QSO PKS1245-19 is also included in a sample of the most powerful radio 
sources~\cite{brl} from the Molonglo catalog \cite{mol}. Molonglo is a 408-MHz survey looking for sources with flux
densities above 0.7~Jy.  This catalog covers
7.85 sr of the sky, over declinations $+18^\circ.5 \geq \delta \leq -85^\circ.0$
and $|b| \geq 3^\circ$.   No other powerful radio sources are found in this catalog in the direction of the HEA\footnote{If we 
modify our search to look for objects within $z < 0.024$, we find
19 sources which fulfill our initial requirement of at least 2~Jy at
a frequency of 2.7~GHz.
From these
only one has an angular distance $\alpha$ of less than $15^\circ$ from the HEA.It is the galaxy pair VV201~\cite{vv201} at $z = 0.015$ or a distance of about 61~Mpc, and $\alpha = 8.5^\circ$ from the event
direction. If a 1~Mpc magnetic field coherence length is assumed, an rather large
extragalactic 
magnetic field of $\sim 4$~nG is necessary to bend a proton to the event 
direction ~\cite{augrep}. The radio luminosity of this pair of interacting galaxies is 
$\sim 5\times 10^{41}$~erg/s, much lower than the $10^{45}$~erg/s required for UHECR sources. This source does not fulfill the $S_\mathrm{min}$ requirement, Eq.~\ref{E:Smin}.}.  PKS1245-19 is also among 
96 sources selected as good candidates for having high Faraday Rotation 
\cite{hifr}.
It has also been listed as an energetic gamma-ray source detected by the Egret 
telescope~\cite{egret}.

This source candidate is therefore very much like the one found in the Fly's Eye
event direction (QSO 3C147). 
An integration of the PKS1245-19 radio spectrum (equation~\ref{eq:lum}) gives a radio 
luminosity of $\sim 3 \times 10^{46}$~erg/s. As both QSOs are also listed as having
high Faraday Rotation measurement, they are excellent candidates for accelerating
particles to ultra high energies. Just as the case with 3C147, PKS1245-19 is very 
far away, at a redshift of 1.275 (3.8~Gpc for current cosmological constants).

\section{AGASA High Energy Events}
\label{sec:agasa}

The Akeno Giant Air Shower Array (AGASA) has detected 11 events with
energies above $10^{20}$~eV. Although there is a $\sim 30$\%
discrepancy among their energy measurement and HiRes'~\cite{demarco},
we take their results as published \cite{agasa}.  Given these
 uncertainties we search for far away sources correlating with the
 AGASA events for completeness. There are other works searching for
 sources of AGASA events \cite{siglag,gorbtrois}  which find no significant correlation.

Following our prescription we search for 
very high luminosity (above $10^{45}$~erg/s) objects in the direction of any of
these 11 events. The only difference is that due to the poorer AGASA
angular resolution, we look for a high luminosity source within a $3^\circ$
angular distance from each event direction. This will of course increase the chance probability
of having an incorrect source in the direction of an event, if sources positions are sampled from an isotropic distribution. 
The AGASA latitude is $35^\circ47'$ and its field
of view covers approximately between $-10^\circ < \delta < 80^\circ$ in Equatorial
coordinates. Our final catalogue of high luminosity
sources in the AGASA field of view consists of 54 objects. From these
the QSO 3C380 is at an angular distance of $2.5^\circ$ from AGASA's 
$1.05 \times 10^{20}$~eV event. This event has $\alpha = 298.5^\circ$ and
$\delta = 18.7^\circ$. The chance probability for a $3^\circ$ correlation from an
isotropic distribution  for one isolated event correlation is of 0.048,
and accounting for the fact that  one event among 11 correlates makes
this probability equal to 0.42.

Estimating the radio luminosity of this quasar from Equation~\ref{eq:lum}, one gets 
$\sim 10^{46}$~erg/s. It is also
listed in the same large Faraday Rotation list~\cite{hifr} as the above
QSOs. Just as the previous candidate sources, QSO 3C380 is an excellent
source for acceleration of UHECR. It is also faraway, at a redshift of 
0.692 (2.4 Gpc for current cosmological parameters).

Since the chance probability for  source correlation with AGASA's
events directions is very large, we might conclude that there is no
  significant correlation with high luminosity objects. However as
  discussed in the introduction there are uncertainties on the
  energies of these events and they might not fulfill our requirement
  of  $10^{20}$˜eV minimum energy.

\section{Exotic candidates}
\label{sec:cand}

As described in the previous sections, QSOs PKS1245-19, 3C147 and 
3C380 whose positions are well-correlated with the directions of Auger's, Fly's Eye's highest energy events, 
and to one of AGASA's high energy events, are very
good UHECR source candidates. The only question then relates to their large distance from
the Earth.

One possibility, that involves new physics but not necessarily new particles, is that 
the propagation distance of UHECR can be increased if Lorentz symmetry is violated \cite{stecker}.
In this work we analyze possible new particles that could be responsible for the high energy
events.

Below we describe two candidates that could account for this huge
distance. One important remark is that although the GZK feature
has been observed, high energy exotic events can certainly compose a small fraction
of the UHECR spectrum. 
They could maybe even account for most of the composition at the highest end of the
spectrum above $10^{20}$~eV. 

Two possible candidates would be axions-like particles~\cite{axi} and exotic massive 
hadrons~\cite{cfk,afk}. Photons would convert into axions at the source,
due to their interaction with the source's magnetic field. They
would then propagate through intergalactic space without scattering on the CMB or other extragalactic background photons, and convert back to photons in the 
Milky Way.  Their air shower would thus be initiated by a 
photon. However, as we will describe below, the photon-axion conversion
probability is low at such high energies.
Alternatively, exotic massive hadrons would be the primary particle in the atmospheric
shower development \cite{afk}. These showers can be 
distinguished from showers initiated by standard particles~\cite{wash}. However, as we will
describe later, it is not clear how these particles can be produced without being
constrained by the diffuse photon flux, since a large number of photons should be produced
simultaneously. 

\subsection{Axion-Like Particles}
Axion-like particles (ALP) ~\cite{axi} would be excellent candidates to propagate through cosmological distances without scattering. They might also explain the possibly anomalous transparency of the universe to TeV gamma rays observed in air Cherenkov telescopes~\cite{Hooper:2007bq,serp}.  In models of photon-axion mixing, an UHE gamma is produced as a byproduct of acceleration of hadronic cosmic rays in the astrophysical source.  This gamma coherently interacts with background magnetic fields and is converted into a UHE axion which can then propagate cosmological distances without scattering.  Upon entering the magnetic field of the Milky Way, the axion reconverts into a UHE gamma which induces a detectable cosmic ray air shower.  Due to low statistics at the highest energies, experimental constraints on a possible gamma-induced fraction of observed air showers are still rather weak, with upper bounds ranging from the $20\%-60\%$ level for UHECR of energies greater than a few $10^{19}$~eV (see for example \cite{Risse:2007}).  It has also been shown that the longitudinal profile of the most energetic Fly's Eye event is compatible with that expected of a photon-initiated air shower at the $1.5\sigma$ level~\cite{Risse:2004mx}.  Morever, due to the poorer composition sensitivity in UHECR detected by surface detectors~\cite{Risse:2005jr,Aglietta:2007yx}, it is not possible to unambigously distinguish between hadronic or photon-initated showers for either the Auger event or the AGASA events.  In particular, there is no composition-sensitive muon flux information on the correlated AGASA event.  So it is possible for all three of the correlated events discussed here to be photon-initiated.

The interaction term 
between the axion field $a$ and photons is
\begin{equation}
\mathcal{L}_\mathrm{int} = -\frac{1}{4} g a F\tilde{F} = g \vec{E}\cdot\vec{B}
\end{equation} 
where $g$ is the axion-photon coupling, $F$ represents the photon field, and
$\vec{E}$ and $\vec{B}$ represent the electric and magnetic fields 
respectively.
In the presence of a background magnetic field of strength $B$, this interaction term 
becomes an off-diagonal term in the effective flavor mixing matrix, coupling axions to
the photon polarization component aligned with the magnetic field 
direction. 
At sufficiently high photon energies $\omega > (m_a^2-m_\gamma^2)/(2 g B)$, the photon-axion system is in a regime of maximal mixing, and the probability of conversion from photons to axions and vice-versa is $P\sim (g B L)^2/4$.  This probability becomes of order unity in regions where $B L \sim 1/g$.   

The UHE photons are expected to be produced as secondary particles in the accleration of UHE protons.  In order to accelerate cosmic ray protons to $E_{\rm proton}=10^{20}$~eV energies, the 
sources must have magnetic field regions which satisfy $B_0 L = E_\mathrm{proton}$.
This coincides with the results of galactic magnetic field models 
\cite{bari} which arrive to a comparable magnetic baseline for the 
poloidal magnetic field, 
$B_0 L \sim (6 \mu\mbox{G})\cdot (4\mbox{ kpc}) = 3\times 10^{19}\mbox{ eV}$.
So an axion model with $g\sim 10^{-11}\mbox{ GeV}^{-1}$ would predict 
efficient conversions via oscillation both at the source, and in the galaxy 
\cite{Hooper:2007bq,serp}.  Alternatively conversions could occur in propagation over sufficiently large coherent domains of the weaker extragalactic magnetic fields \cite{DeAngelis:2007dy}.

However when considering photon-axion oscillations at ultra high energies,
effects from magnetic birefringence might dominate implying in an insignificant
oscillation probability \cite{gunt}. At energies $E >> \mbox{ TeV}$, 
magnetic birefringence due to quantum electrodynamics alters the photon 
dispersion relationship and induces a large diagonal component into the 
effective mixing matrix. As a result the effective photon-axion mixing angle 
becomes too small to produce efficient photon-axion conversion.  A previous paper proposing photon-axion mixing as the source of the correlations between HiRes cosmic ray events and distant BL Lac objects \cite{Abbasi:2005qy} did not account for the QED birefringence effect \cite{Fairbairn:2009zi}, so here we will investigate the effect in detail.

The formulation of the birefringence of the vacuum, was first quoted by 
Adler~\cite{adler}, following numerical evaluations by Toll~\cite{toll}. 
In the presence of a constant background magnetic field $B$, the indices of refraction for photon polarizations perpendicular and parallel to the $B$ are given by:

\begin{equation}
\label{E:Adler}
n_{\bot,\|} = 1 + \frac{\alpha}{\pi}\left(\frac{B \sin\theta}{2 B_\mathrm{cr}}\right)^2 N_{\bot,\|}(x)    
\end{equation}
where $\alpha$ is the fine-structure function, $x$ 
parameterizes the center of mass energy as
\begin{equation}
x = \frac{\omega}{2 m}\frac{B \sin\theta}{B_\mathrm{cr}},
\end{equation}
$\theta$ is the angle between the photon propagation and the direction of the background field $B$, and the critical magnetic field is
\begin{equation}
B_\mathrm{cr} = m_e^2/e = 4.41\times 10^{13}~\mathrm{G}.
\end{equation}
The function $N_{\bot,\|}(x)$ is determined by Toll \cite{toll}, by numerically evaluating the dispersion, averaged over a series of absorption resonances in which the external magnetic field absorbs the excess energy-momentum in the process $\gamma \rightarrow e^+ e^-$.  The numbers given by Adler, and subsequently widely quoted, including in  \cite{raff} are~\footnote{Note that we use here the conventional definition of photon polarization using the electric field vector rather than the opposite definition used by both Toll and Adler.}

\begin{eqnarray}
N_\bot &= 8/45,\\ 
N_\| &= 14/45.
\end{eqnarray}

Most subsequent astrophysical photon-axion mixing papers \cite{Chelouche:2008ta,Chelouche:2008ax,SanchezConde:2009wu,Jain:2009hf} primarily concern themselves with mixing at low energies, and therefore do not explicitly quote Adler's admonition that these values only represent a particular asymptotic limit of Toll's calculations when $B\ll B_c$ and $\omega < 2 m_e$.  This is the limit when the available energy in the interaction between the photon and magnetic field is far below the first pair production resonance, a regime of normal dispersion.  In Toll's thesis, it is illustrated that as $x$ increases through the series of pair production resonances, whose effects can be averaged over, the photon enters a regime of anomalous dispersion where the index of refraction decreases through $n=1$ before becoming slightly negative.  In the limit $B\ll B_\mathrm{cr}$, the numerical predictions of the asymptotic behavior at high $x$ far beyond the resonances are
\begin{eqnarray}
N_\bot &= -0.4372 / x^{4/3},\\
N_\| &= -0.6558 / x^{4/3}. \label{E:anomalous}
\end{eqnarray}
This approximation is valid when $x>500$ or $B_\mathrm{mG}\; \omega_{12} > 10^{13}$, where magnetic field is measured in milliGauss, and photon energy 
in units of TeV.  

The axion-photon mixing matrix is given by~\cite{raff}
\begin{equation}
\left[ \omega - i\partial_z - \left( \begin{array}{ccc} \Delta_\bot^\mathrm{QED} & 0 & 0 \\ 0 & \Delta_\mathrm{pl} + \Delta_\|^\mathrm{QED} & \Delta_\mathrm{B} \\ 0 & \Delta_\mathrm{B} & \Delta_a \end{array} \right) \right] \left( \begin{array}{c}  A_\bot \\ A_\| \\ a \end{array} \right) = 0.
\end{equation}
Where $A_\bot$ and $A_\|$ are orthogonal components of the photon field.
The elements of the dispersion matrix are defined and estimated~\cite{gunt} as

\begin{eqnarray}
\Delta_a = \frac{m_a^2}{2\omega} &\simeq 2.5\times 10^{-20}\ m_{\mu\mathrm{eV}}^2 \omega_{12}^{-1}~\mathrm{cm}^{-1}, \label{E:Da} \\
\Delta_\mathrm{B} = \frac{g B}{2} &\simeq 1.7 \times 10^{-21} \ g_{11} B_\mathrm{mG}~\mathrm{cm}^{-1},\\
\Delta_\mathrm{pl} = \frac{\omega_\mathrm{pl}^2}{2\omega} &\simeq 3.5\times 10^{-26} \left(\frac{n_{e}}{10^3~\mathrm{cm}^{-3}}\right) \omega_{12}^{-1}~\mathrm{cm}^{-1},\\
\Delta_\|^\mathrm{QED}  \simeq -\frac{\alpha}{\pi}  \left(\frac{B}{2 B_\mathrm{cr}}\right)^2 N_\| \omega & = -1.3\times 10^{-21} \ B_\mathrm{mG}^2 \omega_{12}~\mathrm{cm}^{-1}
\end{eqnarray}
where $m_{\mu\mathrm{eV}} \equiv m_a/\mu\mathrm{eV}$, 
$g_{11} \equiv g \times 10^{11}$~GeV, $\omega_{\rm pl}^2$ is the plasma 
frequency for an electron density $n_e$.
As we have now seen, the last equation holds only for low magnetic field and low energy photons such that $x\ll 10^{-1}$ or $B_\mathrm{mG} \cdot \omega_{12} \ll 10^{10}$.

Since the sign of the QED birefringence term $\Delta_\|^\mathrm{QED}$ is negative, indicating a regime of normal dispersion, and $\Delta_\mathrm{pl}$ is negligible, there is no possibility to achieve the resonance condition $\Delta_\|^\mathrm{QED} = \Delta_a$ where the mixing angle $\theta = (1/2) \arctan(2\Delta_\mathrm{B}/(\Delta_\mathrm{QED}+\Delta_\mathrm{pl}-\Delta_a))$ becomes maximal.  The best that can be hoped for is that both $\Delta_\|^\mathrm{QED} , \Delta_a \ll \Delta_\mathrm{B}$ so the configuration is sufficiently close to resonance that a large mixing angle can still be achieved.  Suppose that the initial conversion from photons to axions occurs in the galaxy hosting the astrophysical photon source, and the regeneration of photons occurs in the Milky Way galaxy.  Then, the highest energy photons that can penetrate the wall of extragalactic background light can be estimated using $g_{11} = 10$, $B_\mathrm{mG} = 10^{-4}$, and $L \sim 30$~kpc for both source and destination galaxy so that the mixing probability is of order unity in each.  In this case, the maximum photon energy for which this process can occur efficiently is $10^{17}$~eV, for which $\Delta_a \sim 10^{-25}~\mathrm{cm}^{-1}$, $\Delta_\|^\mathrm{QED} \sim \Delta_\mathrm{B} \sim 10^{-24}~\mathrm{cm}^{-1}$.  

The only remaining possibility to obtain efficient mixing at even higher energies is to enter the regime of anomalous QED dispersion, in which the indices of refraction are parameterized by Eqs.~\ref{E:anomalous}.  In this case, $\Delta_\|^\mathrm{QED}>0$, and so an exact resonance is possible.   To achieve this condition, the center of mass energy must be well above the energy required to produce on-shell pairs, $B_\mathrm{mG} \cdot \omega_{12} > 10^{13}$ \cite{toll}.  For photon energies of $10^{19}$~eV, the magnetic field required is then $10^{4}$~G.  Such high fields may indeed be present in the high energy photon sources, and induce resonant conversion of photons into axions, or even efficient conversion by adiabatic level-crossing.  However, there are no such fields in the Milky Way, and the resulting ultra-high-energy axions cannot be efficiently reconverted into detectable photons. 

\subsection{Exotic Massive Hadrons}
Before the high energy end of the cosmic ray spectrum was established,
a class of exotic massive hadrons was proposed \cite{cfk}
to account for the events beyond the GZK cutoff. This generic
class was coined as ``uhecron'', and is composed of colored,
strongly interacting massive particles. 

Uhecrons are modeled \cite{afk} as a heavy single constituted core surrounded by light 
hadronic degrees of freedom. The core concentrates the momentum of the particle while the
light degrees of freedom (which can be gluons and/or light quarks) is responsible for the
interaction. The strong interaction is necessary in order to produce a hadronic
shower in the atmosphere \cite{afk}.  If the uhecron is too heavy, 
the momentum carried by the light consituent will be too small since most of the energy 
is carried by its core and no interaction in the air
will occur. It was shown \cite{afk} that in order to produce a hadronic shower in the
atmosphere, an uhecron cannot be heavier than 50~GeV.
It is important to note that particles such as the proposed S$^o$ \cite{cfk} cannot be classified 
as an uhecron, since its momentum is shared almost equally by all its constituents.

Uhecrons can avoid the GZK cutoff since it loses much less energy than
protons while traveling through the CMB. In short, their heavy mass shifts the threshold for 
photoproduction of pions to much higher energies. They can propagate distances of gigaparsecs 
through the CMB before losing most of its energy~\cite{cfk}. Uhecrons are also taken as 
electrically neutral particles in order to avoid deflections by magnetic fields.

Simulations of uhecron induced air showers \cite{afk} show that
uhecrons with masses up to 50~GeV are compatible with UHECR
events. The shower generated by the interaction of an uhecron with
the atmosphere, can resemble the longitudinal profile of the highest
energy cosmic ray detected by Fly's Eye \cite{fly}.

One of the best candidates for an uhecron is the heavy gluino lightest
supersymmetric particle (LSP), proposed by Stuart Raby~\cite{raby}. It
fits all the uhecron requirements. Experimental limits set the heavy gluino
LSP mass between 25 and 35~GeV~\cite{hgexp}.

As uhecrons can account for the large distances between faraway QSOs  
and the Earth, the heavy gluino
LSP is a candidate for the highest energy cosmic rays analyzed here.

It has also been showed that it is possible to distinguish uhecron
from proton or heavy nuclei induced showers \cite{wash}. Uhecrons
with masses in the 25 to 35~GeV window can be discriminated from
the bulk of the cosmic ray spectrum,which might be composed by protons
or iron. An enhancement of this discrimination can be achieved by
correlating the event direction with the far away source.

However, the acceleration mechanism to produce uhecrons at these energies
is not well established. It has been pictured generically \cite{cfk} as
originating from proton collisions with hadronic matter in the source.
This requires protons being accelerated to energies slightly above 
$10^{20}$~eV. It was pointed out \cite{cfk} that even if uhecrons are
originated in proton collisions, it is hard to explain the lack of lower
energies uhecrons. This remark does not hold here, since uhecrons
would constitute a small fraction of the UHECR spectrum, and the lower
energy ones would be hidden by the larger fraction of standard particles.
Information from the Large Hadron Collider on the heavy gluino LSP will
contribute to the possibility of uhecrons accounting for the long distances
between QSOs and Earth.  

The main potential problem with uhecron production from higher energy proton
collisions is that since the uhecron is modeled as a neutral particle,
photons will also be produced in abundance. If all similar sources as the ones found
both in the Auger and Fly's Eye event direction, produce uhecrons, a significant
flux of photons should be found. As the Fermi measurement on the diffuse extragalactic gamma
flux \cite{fermi} is even lower than the one from Egret, it seems that most top-down scenarios are
excluded as significant contribution for the UHECR spectrum \cite{guntertopd}.

\section{Conclusions}

We have described our search for the source of the highest energy cosmic ray 
detected by the Pierre Auger Observatory. We show that given requirements
on the acceleration capabilities of the source, one faraway quasar
is located at $1.2^\circ$ from the event arrival direction. The chance probability
for this correlation from an isotropic distribution is of 0.013.

In this search we look for sources with radio luminosities above 
$10^{45}$~erg/s, which are required in order to be able to accelerate
particles to energies of $10^{20}$~eV.  This requirement is valid for
many acceleration models, as the diffusive or Fermi mechanisms, but
not for some mechanisms as the inductive acceleration close to central
black holes in AGNs. Also although the bound is set for magnetic
luminosities we use it as a bound for bolometric luminosities as in \cite{martin}. The quasar PKS1245-19 found
in the direction of the Auger event, has a luminosity of 
$4 \times 10^{46}$~erg/s. It is included in a catalogue of the most powerful
radio sources \cite{mol} and among 96 sources selected to be good candidates
for having large Faraday Rotation \cite{hifr}. It is also listed as
an energetic gamma-ray source by the Egret telescope \cite{egret}.
PKS1245-19 is therefore very much like quasar 3C147 found~\cite{som} at one 
sigma from the direction of the highest energy cosmic ray ever 
detected~\cite{fly}.

As no obvious astrophysical source within the GZK cutoff was found in the direction of 
either the Auger or the Fly's Eye event, the faraway
quasars are so far the best observed candidate sources for these high energy cosmic rays. 
In this case
the only explanation for these particles being able to reach us, is that
they are composed of non-standard particles, at least during their transit,
or a non standard propagation mechanism.

Two plausible candidates are axions and exotic massive hadrons. As was
shown in this work, the probability for photon-axion conversion
at these high energies is probably too small for this effect to produce a significant contribution to the UHECR flux. Uhecrons could account for these events, but there are potential problems
related to their origin. 
Accelerating protons to even higher energies above $10^{20}$~eV is required
in order to originate uhecrons at the required energy. This mechanism will simultaneously
generate high energy photons, which flux is limited by current gamma telescopes.
One should however
bear in mind that the acceleration mechanism for standard particles is not well
understood either.  Furthermore, it could be that the energies of these exotic uhecron-induced showers are systematically overestimated since the energy scale of the various detectors is calibrated to the average, presumably hadronic, cosmic ray.  If either the shower profile or the invisible energy carried by neutral shower particles differs from that expected from typical hadronic showers, then the energy measurement can be biased~\cite{Chou:2006jj} for a small fraction of the cosmic ray flux composed of exotic particles.  

As we are analysing events at energies well beyond that produced
in Earth based accelerators, there could still be some
unknown exotic phenomena involved in the cosmic ray production and propagation.  
It has been pointed out that violation of Lorentz symmetry could greatly increase the propagation distance of UHECR by changing the kinematics of the scattering on background photons \cite{stecker}. 
Although we cannot find non problematic candidates that could account for these events, it is 
clear that there exist faraway quasars 
in the direction of these ultra high energy events and that only new
physics can account for a particle propagating from their distance to
us.  The alternative of course, is that UHECR sources are nearby, but
transient and not readily apparent in current observations of the
sky. Or that they are heavier nuclei which do not point to their
sources.

We thank the members of the Auger collaboration at Fermilab, for valuable
discussions. AC thanks Steve Adler for graciously providing a copy of
Toll's thesis from the Princeton library. IA was partially funded by
the U.S.~Department 
of Energy under contract number DE-AC02-07CH11359 and the Brazilian National 
Counsel for Scientific Research (CNPq).  AC is also supported by the U.S. Department of Energy under contract No. DE-AC02-07CH11359.
This work made use of the NASA/IPAC Extragalactic Database (NED) 
which is operated by the Jet Propulsion Laboratory, California Institute of 
Technology, under contract with the National Aeronautics and Space 
Administration.


\begin{thebibliography}{99}
\bibitem{augani} J.~Abraham {\it et al.},
Astropart.\ Phys.\  {\bf 29}, 188 (2008)
[Erratum-ibid.\  {\bf 30}, 45 (2008)];
J.~Abraham {\it et al.},
Science {\bf 318}, 938 (2007).
\bibitem{hragn} R.~U.~Abbasi {\it et al.},
Astropart.\ Phys.\  {\bf 30}, 175 (2008).
\bibitem{hrgzk} R.~Abbasi {\it et al.},
Phys.\ Rev.\ Lett.\  {\bf 100}, 101101 (2008).
\bibitem{augzk} J.~Abraham {\it et al.},
Phys.\ Rev.\ Lett.\  {\bf 101}, 061101 (2008)
\bibitem{gzk} K.~Greisen, Phys. Rev. Lett. {\bf 16}, 748 (1966);
G.~T. Zatsepin, V.~A. Kuz'min, JETP Lett. {\bf 4}, 78 (1966);
ZhETF Pis'ma Eksp {\bf 4}, 114 (1966).

\bibitem{Waxman:2004ez}
E.~Waxman,
Nucl.\ Phys.\ Proc.\ Suppl.\  {\bf 151}, 46 (2006)
[arXiv:astro-ph/0412554].
\bibitem{Lemoine:2009pw}
M.~Lemoine and E.~Waxman,
JCAP {\bf 0911}, 009 (2009)
[arXiv:0907.1354 [astro-ph.HE]].
\bibitem{Farrar:2008ex}
G.~R.~Farrar and A.~Gruzinov,
Astrophys.\ J.\  {\bf 693}, 329 (2009)
[arXiv:0802.1074 [astro-ph]].
\bibitem{Waxman:2008bj}
E.~Waxman and A.~Loeb,
JCAP {\bf 0908}, 026 (2009)
[arXiv:0809.3788 [astro-ph]].
\bibitem{som} J.~W.~Elbert and P.~Sommers,
Astrophys.\ J.\  {\bf 441}, 151 (1995).
\bibitem{fly} D.~J.~Bird {\it et al.},
Astrophys.\ J.\  {\bf 441}, 144 (1995).
\bibitem{vc} M.-P.~V\'{e}ron-Cetty and P.~V\'{e}ron,
Astron.~\&~Astrophys.\ {\bf 455}, 773 (2006).  
\bibitem{demarco} D.~De Marco, P.~Blasi and A.~V.~Olinto,
Astropart.\ Phys.\  {\bf 20}, 53 (2003).
\bibitem{agasa} M.~Takeda {\it et al.},
Astrophys.\ J.\  {\bf 522}, 225 (1999)
\bibitem{siglag} G.~Sigl, D.~F.~Torres, L.~A.~Anchordoqui and G.~E.~Romero,
  Phys.\ Rev.\  D {\bf 63}, 081302 (2001)
  [arXiv:astro-ph/0008363].
\bibitem{gorbtrois} D.~S.~Gorbunov and S.~V.~Troitsky,
  Astropart.\ Phys.\  {\bf 23}, 175 (2005)
  [arXiv:astro-ph/0410741].
\bibitem{Abbasi:2005qy}
R.~U.~Abbasi {\it et al.}  [HiRes Collaboration],
Astrophys.\ J.\  {\bf 636}, 680 (2006)
[arXiv:astro-ph/0507120].
\bibitem{Gorb} D.~S.~Gorbunov, P.~G.~Tinyakov, I.~I.~Tkachev and S.~V.~Troitsky,
  JETP Lett.\  {\bf 80}, 145 (2004)
  [Pisma Zh.\ Eksp.\ Teor.\ Fiz.\  {\bf 80}, 167 (2004)]
  [arXiv:astro-ph/0406654].
\bibitem{hillas} A.~M.~Hillas,
Ann.\ Rev.\ Astron.\ Astrophys.\  {\bf 22}, 425 (1984).
\bibitem{herbig} T.~Herbig and C.~S.~Readhead,
The Astrophys.\ Journ.\ Supp.\ Series \ {\bf 81}, 83 (1992).
\bibitem{hifr} M.~Inoue, H.~Tabara, T.~Kato and K.~Aizu,
Publ. Astron. Soc. Japan {\bf 47}, 725 (1995)
\bibitem{kato} T.~Kato, H.~Tabara, M.~Inoue and K.~Aizu,
Nature\ {\bf 341}, 720 (1989).
\bibitem{martin} M.~Lemoine and E.~Waxman,
JCAP {\bf 0911}, 009 (2009).
\bibitem{wallp} J.~V.~Wall and J.~A.~Peacock,
Mon.~Not.~R.~astr.~Soc. {\bf 216}, 173 (1985).
\bibitem{ned} Mazzarella, J.~M., \& The NED Team, Astronomical Data 
Analysis Software and Systems XVI, {\bf 376}, 153 (2007); 
Proceedings Edited by Richard A. Shaw, Frank Hill and David J. Bell,
Conference 15-18 October 2006, Tucson, Arizona, USA.
\bibitem{brl} P.~R.~Best, H.~J.~A.~Rottering and M.~D.~Lehnert,
Mon.~Not.~R.~astr.~Soc. {\bf 310}, 223 (1999).
\bibitem{mol} M.~I.~Large {\em et el.}, Mon.~Not.~R.~astr.~Soc. {\bf 194}, 693 
(1981).
\bibitem{vv201} Vorontsov-Velyaminov, 
Astr. Ap. Suppl. {\bf 28}, 1 (1977).
\bibitem{augrep} Auger Collaboration,
``The Pierre Auger Project Design Report,''
FERMILAB-PUB-96-024 (1996).
\bibitem{egret} C.~E.~Fichtel {\em et el.}, The Astrophysical Journal Supplement 
Series, {\bf 94}, 551 (1994).
\bibitem{stecker} S.~R.~Coleman and S.~L.~Glashow,
  Phys.\ Rev.\  D {\bf 59}, 116008 (1999)
  [arXiv:hep-ph/9812418];
S.~L.~Dubovsky and P.~G.~Tinyakov,
  Astropart.\ Phys.\  {\bf 18}, 89 (2002)
  [arXiv:astro-ph/0106472];
F.~L.~Stecker, New Journal of Physics {\bf 11}, 085003 (2009).
\bibitem{axi} S.~J.~Asztalos {\em et el.},
Ann.\ Rev.\ Nucl.\ Part.\ Sci.\  {\bf 56}, 293 (2006).
\bibitem{cfk} D.~J.~H.~Chung, G.~R.~Farrar and E.~W.~Kolb,
Phys.\ Rev.\  D {\bf 57}, 4606 (1998).
\bibitem{afk} I.~F.~M.~Albuquerque, G.~R.~Farrar and E.~W.~Kolb,
Phys.\ Rev.\  D {\bf 59}, 015021 (1999).
\bibitem{wash} I.~F.~M.~Albuquerque and W.~R.~Carvalho,
Phys.\ Rev.\  D {\bf 80}, 023006 (2009).
\bibitem{bari} D.~Harari, S.~Mollerach and E.~Roulet,
JHEP {\bf 9908}, 022 (1999).
\bibitem{Hooper:2007bq}
D.~Hooper and P.~D.~Serpico,
Phys.\ Rev.\ Lett.\  {\bf 99}, 231102 (2007)
\bibitem{serp} M.~Simet, D.~Hooper and P.~D.~Serpico,
Phys.\ Rev.\  D {\bf 77}, 063001 (2008).
\bibitem{Risse:2005jr}
M.~Risse {\it et al.},
Phys.\ Rev.\ Lett.\  {\bf 95}, 171102 (2005)
[arXiv:astro-ph/0502418].
\bibitem{Aglietta:2007yx}
J.~Abraham {\it et al.}  [Pierre Auger Collaboration],
Astropart.\ Phys.\  {\bf 29}, 243 (2008)
[arXiv:0712.1147 [astro-ph]].
\bibitem{Risse:2007} M.~Risse and P.~Homola,
  Mod.\ Phys.\ Lett.\  A {\bf 22}, 749 (2007)
  [arXiv:astro-ph/0702632].
\bibitem{Risse:2004mx}
M.~Risse, P.~Homola, D.~Gora, J.~Pekala, B.~Wilczynska and H.~Wilczynski,
Astropart.\ Phys.\  {\bf 21}, 479 (2004)
[arXiv:astro-ph/0401629].
\bibitem{DeAngelis:2007dy}
A.~De Angelis, O.~Mansutti and M.~Roncadelli,
Phys.\ Rev.\  D {\bf 76}, 121301 (2007)
[arXiv:0707.4312 [astro-ph]].
\bibitem{gunt} K.~A.~Hochmuth and G.~Sigl,
Phys.\ Rev.\  D {\bf 76}, 123011 (2007).

\bibitem{Fairbairn:2009zi}
M.~Fairbairn, T.~Rashba and S.~Troitsky,
arXiv:0901.4085 [astro-ph.HE].
\bibitem{adler} S.~L.~Adler,
Annals Phys.\  {\bf 67}, 599 (1971).
\bibitem{toll} J.~S.~Toll,
PhD. Thesis, Princeton University (1952).
\bibitem{raff} G.~Raffelt and L.~Stodolsky,
Phys.\ Rev.\  D {\bf 37}, 1237 (1988).
\bibitem{Chelouche:2008ta}
D.~Chelouche, R.~Rabadan, S.~Pavlov and F.~Castejon,
Astrophys.\ J.\ Suppl.\  {\bf 180}, 1 (2009)
[arXiv:0806.0411 [astro-ph]].
\bibitem{Chelouche:2008ax}
D.~Chelouche and E.~I.~Guendelman,
Astrophys.\ J.\  {\bf 699}, L5 (2009)
[arXiv:0810.3002 [astro-ph]].
\bibitem{SanchezConde:2009wu}
M.~A.~Sanchez-Conde, D.~Paneque, E.~Bloom, F.~Prada and A.~Dominguez,
Phys.\ Rev.\  D {\bf 79}, 123511 (2009)
[arXiv:0905.3270 [astro-ph.CO]].
\bibitem{Jain:2009hf}
P.~Jain and S.~Mandal,
arXiv:0910.3036 [astro-ph.CO].
\bibitem{raby} S.~Raby,
Phys.\ Lett.\  B {\bf 422}, 158 (1998).
\bibitem{hgexp}
A.~Mafi and S.~Raby,
Phys.\ Rev.\  D {\bf 62}, 035003 (2000).
\bibitem{fermi} A.~A.~Abdo {\it et al.}  [The Fermi-LAT collaboration],
  Phys.\ Rev.\ Lett.\  {\bf 104}, 101101 (2010)
  [arXiv:1002.3603 [astro-ph.HE]].
\bibitem{guntertopd}D.~V.~Semikoz and G.~Sigl,
  JCAP {\bf 0404}, 003 (2004)
  [arXiv:hep-ph/0309328].
\bibitem{Chou:2006jj}
A.~S.~Chou,
Phys.\ Rev.\  D {\bf 74}, 103001 (2006)
[arXiv:astro-ph/0606742].

\end{thebibliography}
\end{document}